# High performance WR-1.5 corrugated horn based on stacked rings

Bruno Maffei*[a], Arndt von Bieren[b], Emile de Rijk[b], Jean-Philippe Ansermet[c], Giampaolo Pisano[a], Stephen Legg[a], Alessandro Macor[b]

[a]JBCA, School of Physics and Astronomy, The University of Manchester, Manchester M13 9PL, UK; [b]SWISSto12 SA, 1015 Lausanne, Switzerland; [c]Institute of Condensed Matter Physics, EPFL, 1015 Lausanne, Switzerland.

## ABSTRACT

We present the development and characterisation of a high frequency (500 – 750 GHz) corrugated horn based on stacked rings. A previous horn design, based on a Winston profile, has been adapted for the purpose of this manufacturing process without noticeable RF degradation. A subset of experimental results obtained using a vector network analyser are presented and compared to the predicted performance. These first results demonstrate that this technology is suitable for most commercial applications and also astronomical receivers in need of horn arrays at high frequencies.

## 1. INTRODUCTION

Because of their unbeaten RF performance, corrugated waveguides and feedhorns have a large range of applications from communication systems to plasma physics and experimental astrophysics. In particular, most of the radio and mm-wave receivers use such feedhorns to accurately define the beam of their detectors. For instance, Planck instruments include about 50 corrugated horns divided between spectral bands covering a frequency range of 30 to 900 GHz.

Emphasis is now on developing very large focal plane arrays in order to reach high sensitivities. Moreover, each of these horns needs to have a homogeneous performance and low systematic effects. So far these feedhorns were either machined or manufactured through electroforming techniques. For large focal planes of hundreds or thousands of pixels, such costly and time consuming manufacturing techniques cannot be used anymore. Several new techniques are being developed. In the last few years, platelet technology, where horn arrays are made of stacked plates, has been successfully developed for frequencies up to 100 - 150 GHz. Stereolithography is also being investigated with some success for similar frequencies. However, even if silicon micromachining has been used recently, for frequencies above 200 GHz simplified and cheaper mass production of high performance corrugated feedhorns is still an issue.

SWISSto12 has developed the 'stacked rings' technology [1] (patent pending) that has been proven to work reliably for wave-guiding applications in the sub-millimetre and millimetre-wave range and Bragg filters up to 140 GHz and beyond [2]. Recently, a modular wave-guiding system based on this technology has been presented for the WR-1.5 band (500 - 750 GHz), demonstrating the scalability of the technology to frequencies approaching 1 THz. It consists of reconfigurable corrugated waveguide straight sections and bends for flexible use in various applications. For efficient interfacing with conventional rectangular waveguides we have developed a full band WR-1.5 corrugated horn converter including a circular to rectangular waveguide transition that allows for RF characterisation using a vector network analyser (VNA).

The geometry of the horn is based on a previous design developed by JBCA that has been slightly adapted to be more suitable for the stacked rings manufacturing technology. Co- and cross-polarisation beam measurements have been performed across the full band along with a detailed analysis in terms of return loss, insertion loss, directivity, cross-polarisation and radiation pattern. Results show that the measured performance is matching fairly well the expected characteristics from simulations, validating the 'stacked rings' technology for use at frequencies up to at least 750 GHz. In addition, our study shows how this technology opens new possibilities regarding choice of material, geometry and cost-effective applications.

*Bruno.maffei@manchester.ac.uk; phone 44 161 2754141

## 2. CORRUGATED HORN DESIGN

Starting from a previous W-band (75-110 GHz) corrugated horn design with a circular waveguide output, a scaled design has been produced in order to operate within the band 500 – 750 GHz. This involved scaling the dimensions of the horn with the ratio of the frequencies and adapting the shape of the horn to give an aperture diameter of 8mm required for interfacing with the SWISSto12 corrugated wave-guiding system. This gave a beam Full Width Half Maximum (FWHM) of about 5 degrees at the centre of the spectral band. The corrugation parameters and the horn shape have then been modified in order to adapt the design to ease the fabrication of the horn with the ring technology. A rectangular (WR-1.5) to circular transition was then designed and added to the geometry of the horn in order to provide a rectangular waveguide interface at the horn input and to match it to the input of a VNA equipped with frequency extension heads.

### 2.1 Design and modelling of the horn front section

Corrugations

The shape of the horn is based on a modified Winston horn [3] for which corrugations have been included in order to achieve a Gaussian main beam with low sidelobes, a better return loss and a low cross-polarisation [4,5]. This geometry has also been chosen due to the fact that it creates the phase centre right at the aperture of the horn, which, in this specific case, provides a better match to the $HE_{11}$ mode required for low-loss wave-guiding in the rest of the quasi-optical set-up in which the horn is used [6]. The first phase of modelling focussed on optimising the shape and corrugation parameters. In order to efficiently scan a wide range of geometries in this initial phase, simulations were performed using a mode-matching software (CORRUG from SMT Consultancies Ltd), which has the advantage of running fairly fast in comparison to software based on other methods (such as the Finite Element Method).

While in general the corrugation period is usually taken as $\lambda/4$, $\lambda$ being the central wavelength of the spectral band of interest, and each of the tooth width and slot width set to $\lambda/8$, other designs might have different slot and tooth width ratios. In this case we have decided to set these parameters to values that will facilitate and lower the costs of the manufacturing with the ring technology. The thickness of the rings having set values, we chose to have a thinner tooth width and a larger slot width. The tooth width was set to 50 μm (~$\lambda/10$) while the slot width was set to 100 μm (~$\lambda/5$). This choice was also set by the other waveguide component that this horn would need to be used in conjunction with [6].

Linearisation

In order to ease the fabrication of the horn with the stacked rings technology, the shape of the horn has then been modified from a design with curved sections to a design with linearised approximation of the curved sections. This has been achieved by using 11 linear sections along most of the length of the horn except for the waveguide section where the impact on the horn performance would be the highest. We took the opportunity of this design simplification to also shorten the horn, giving a length of 65 mm for the linear version instead of 85 mm for the original horn. Figure 1 gives a comparison of the two geometries. For clarity of the plot the linear version is presented without the corrugations and only the shape is shown in comparison to the original curved corrugated design. Corrugations on the linear version have exactly the same design as for the original horn version.

Figure 2 shows the return loss difference between these two horn geometries. The variation of the return loss across the whole band between the two designs is small. A maximum return loss of -20 dB was judged to be enough for this first prototype (only slightly higher towards the low frequency edge of the band). Further optimisation will allow the reduction of the return loss if necessary in the future.

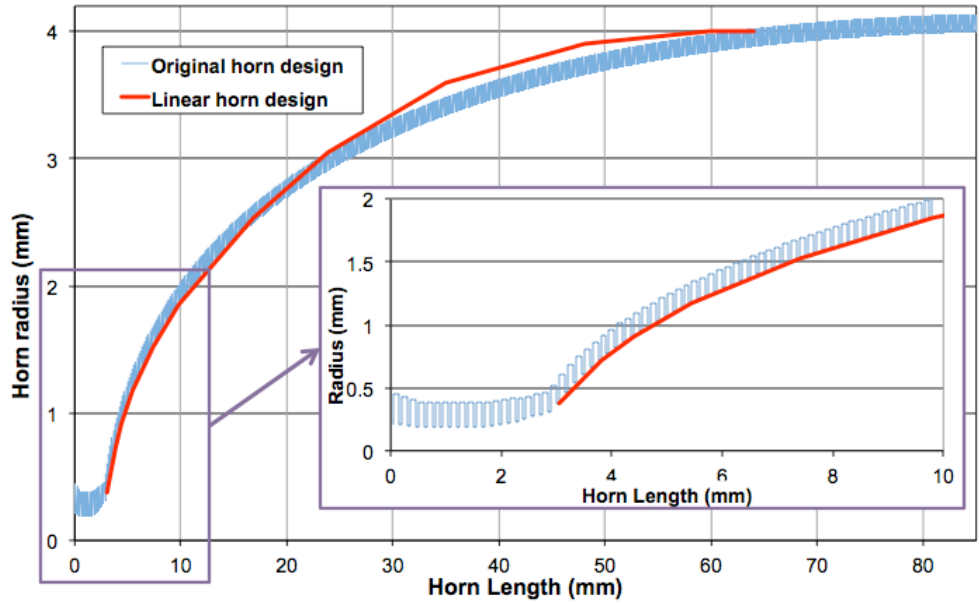

Figure 1. Horn geometry comparison between the original curved version (blue) and the equivalent version approximated with 11 linear sections (red) shown without corrugations. A zoom on the waveguide – horn throat part is also shown.

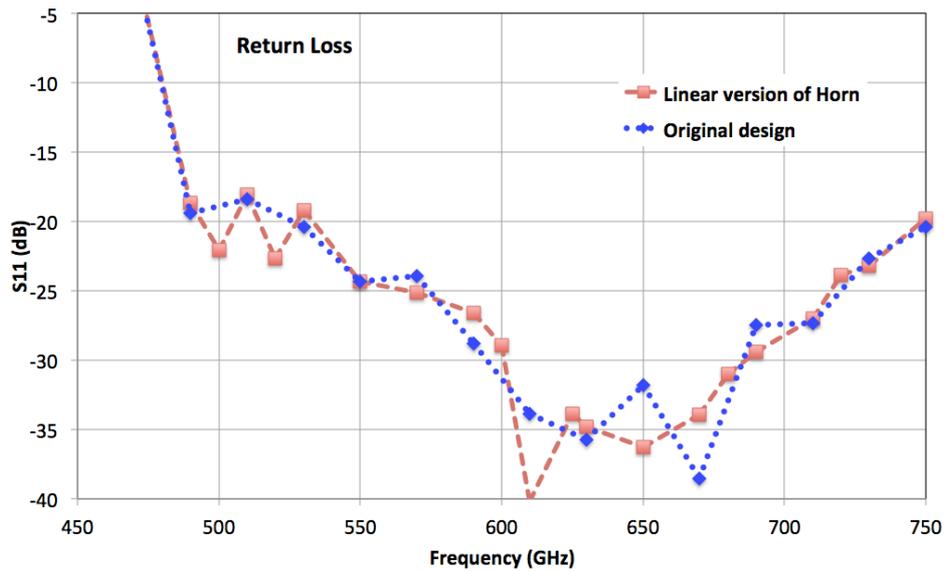

Figure 2. Return loss comparison between the original (dotted blue) and linear (dashed red) versions of the horn.

**2.2 Design and modelling of the circular to rectangular waveguide transition**

The circular to rectangular waveguide transition has been designed using a Finite Element Method (FEM) software (HFSS from Ansoft). The circular waveguide diameter is 0.42 mm while a WR1.5 rectangular waveguide that allows connecting the horn to other rectangular waveguide components has dimensions of 0.381 x 0.1905 mm. In order to have a short device, a stepped transition (Figure 3 left) using a square section and a rectangular intermediate has been designed. The predicted return loss is below -25 dB across the full spectral band (Figure 3 right).

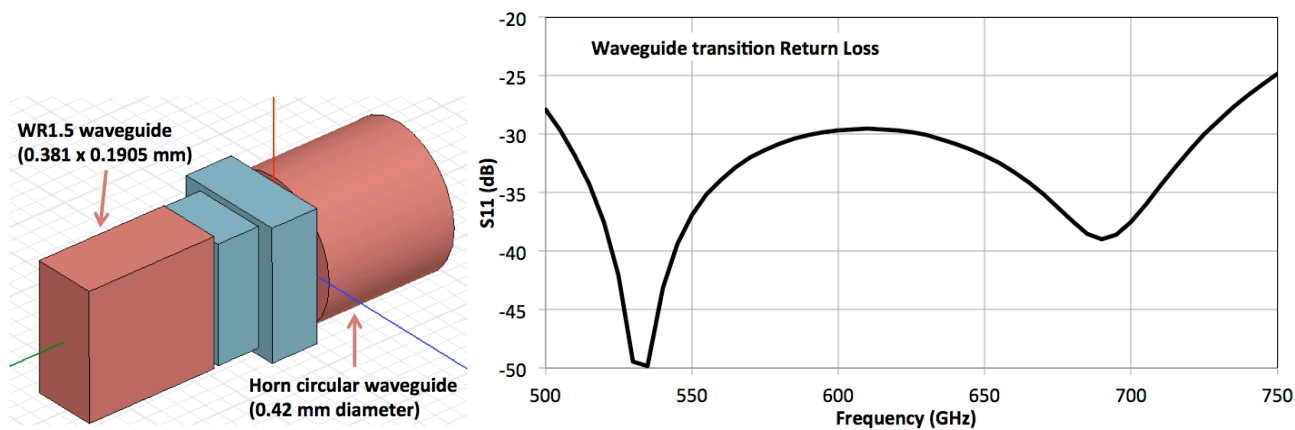

Figure 3. Left: stepped circular to rectangular waveguide transition model. Right: Plot of the stepped transition return loss across the full spectral band modelled with HFSS.

## 2.3 Full horn modelling

The complete horn assembly including the step transition to the rectangular waveguide cannot be modelled using our mode matching code, which is limited to circular symmetric structures. The length (about 130 λ) and aperture (16 λ) diameter of the horn being large comparatively to the wavelength, the FEM-based simulation becomes very demanding in terms of computer RAM and computation time, and so are other methods which do not use approximations. In order to speed up the computation time, models have been run on a chopped version of the full horn to compute the return loss after verification that the results at a few frequencies were similar to the ones run on the full horn model. Calculations have also been performed with two different methods and software for results cross-check. For this purpose, we have chosen to use the Method of Moment (MoM) with the software package FEKO from EM Software & Systems-S.A and FEM with HFSS. Some of these simulation results are presented together with the experimental measurements in Figure 6 (return loss) and Figure 7 (insertion loss).

## 3. MANUFACTURING

The horn consists of a brass guiding tube holding a series of separately machined brass rings in place. Various ring shapes form the step transition, the horn's throat region as well as the taper up to the final diameter of 8 mm. The entire stack of rings is compressed between (i) a 5 mm short rectangular waveguide section integrated in a standard UG-387 waveguide interface and (ii) compression screws that are part of an auto-aligning precision circular waveguide interface on the other side of the horn (see Figure 4). The latter has been designed for fast and reliable connection with other corrugated waveguide components [6].

## 4. RESULTS

The horn has been characterised with an Agilent PNA vector network analyser coupled to WR-1.5 frequency extenders from Virginia Diodes Inc. The return loss is determined by measuring $S_{11}$ with the horn radiating into free space (or onto a foam absorber). The insertion loss has been estimated by shorting the horn using a metallic mirror positioned at its aperture and measuring $S_{11}$ again. This is not a highly accurate way to measure the insertion loss due to the possibility of mismatch between the two components and possible distortions of the propagation inside the horn but gives the worst-case value of the insertion loss.

Co- and cross-polarisation 3D beam measurements have been performed with the horn mounted on the static TxRx extender module on one end, and with a scanning Rx extender module mounted on a 3-axes translation stage at the other end (Figure 5). Varying the distance between the two modules allows for measuring the beams in the near-field and in the far-field.

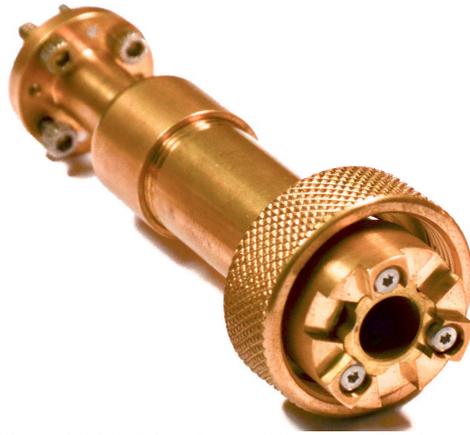

Figure 4. Picture of the manufactured horn, highlighting the precision auto-aligning waveguide flange for interfacing with 8 mm diameter corrugated waveguide components.

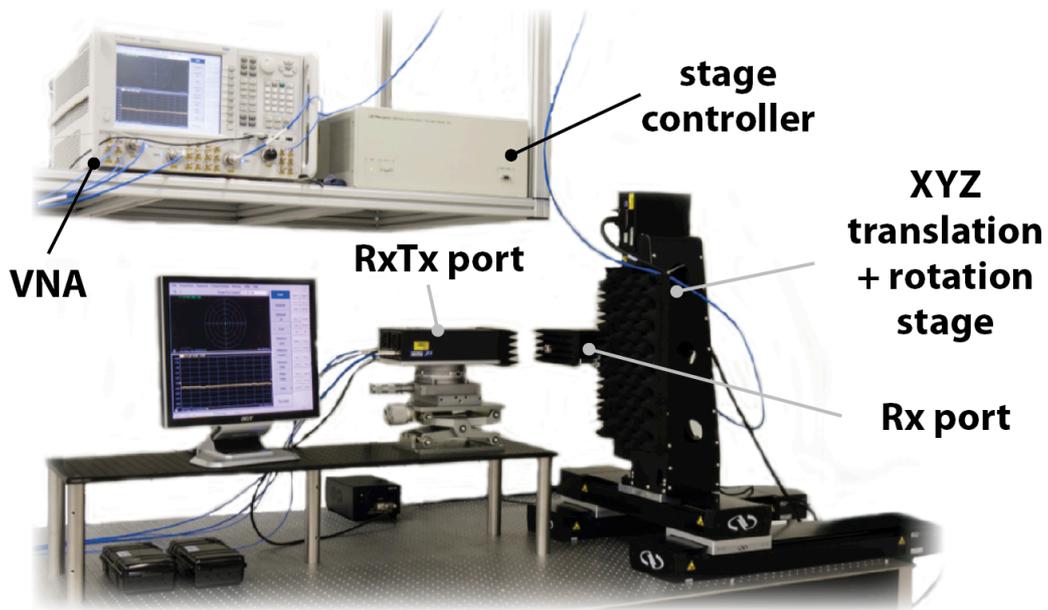

Figure 5. Test set-up for the beam characterisation of the horn using an Agilent PNA coupled to WR-1.5 frequency extenders from Virginia Diodes Inc.

### 4.1 Return loss

The measurements of the return loss are presented in Figure 6 together with various models for comparison: two models of the horn on its own using two methods (MoM and mode-matching) and the model of the horn with the waveguide transition (MoM with FEKO). In order to explain the return loss increase between 600 and 650 GHz when the transition is added in the model, the return loss of the transition on its own is also shown. While some resonance peaks are present, most likely due to a mismatch at the rectangular waveguide interface, the overall trend follows the model (green solid curve). Even though, most of the measured return loss value across the band is below -20dB with an average of -23.2dB

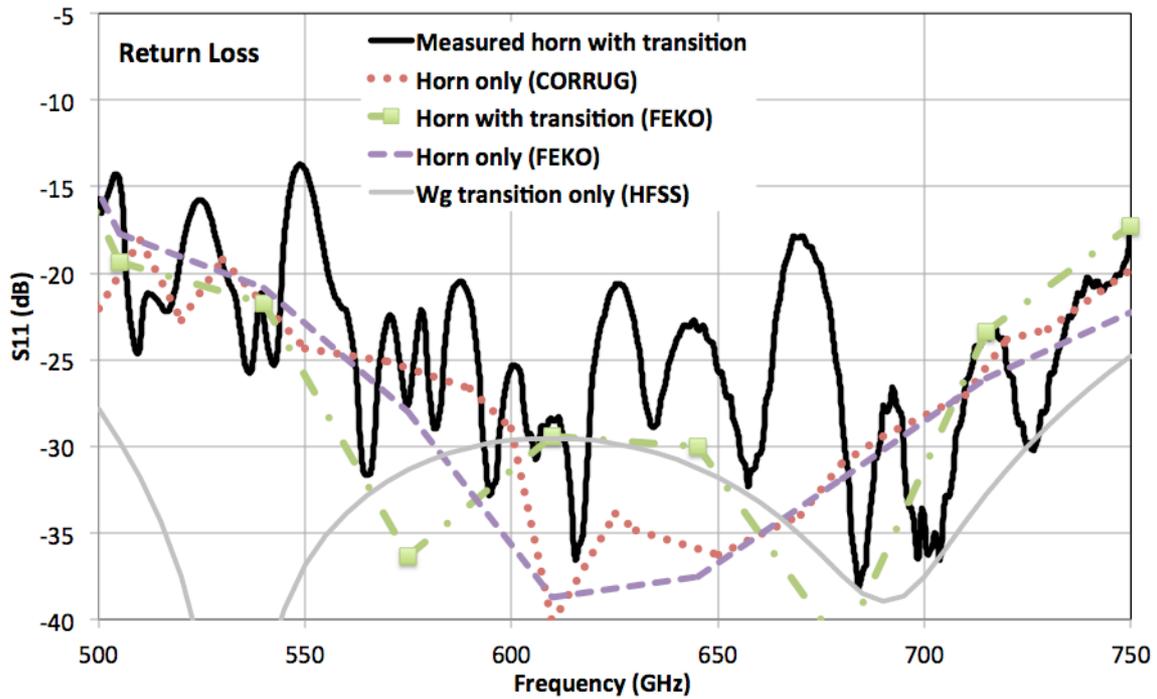

Figure 6. Measured return loss (black solid line, smoothed using a 50-samples moving average filter) compared to 3 models: two of the horn on its own using MoM-FEKO (purple dashed) and mode-matching – CORRUG (brown dotted line) and the model of the horn with the waveguide transition (green solid). The model of the waveguide transition is also added (grey solid) showing its contribution to the full horn return loss increase between 600 and 650 GHz.

### 4.2 Insertion loss

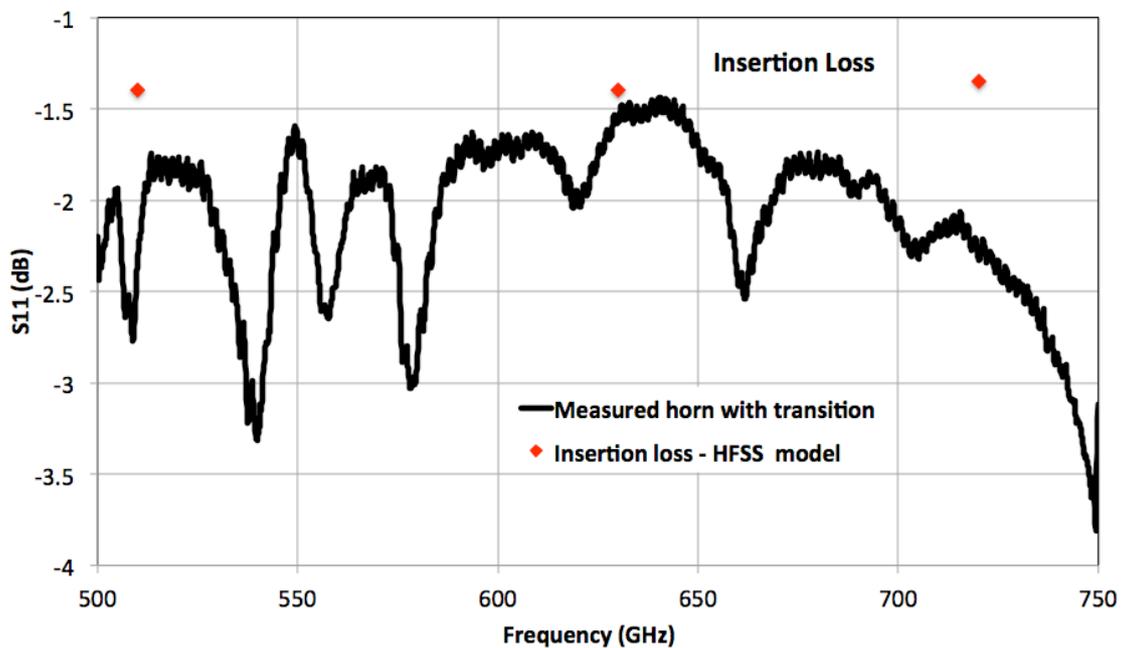

Figure 7. Measured insertion loss

The insertion loss has been calculated using the full size linear horn model for a few frequencies only. As stated above, the measurement technique gives an upper value of the insertion loss but gives however a fair comparison with the model (Figure 7). The higher loss for frequencies above 700 GHz is probably due to a combination of the measurement technique (match between the horn aperture and the mirror) and the fabrication process.

### 4.3 Co- and cross-polarisation beams

Co- and cross-polarisation beams are presented for three frequencies (510, 630 and 720 GHz) across the band in Figures 8, 9 and 10 respectively. Comparison with the model shows that the main beam is well in agreement with the predictions while sidelobes are a few dBs higher than expected.

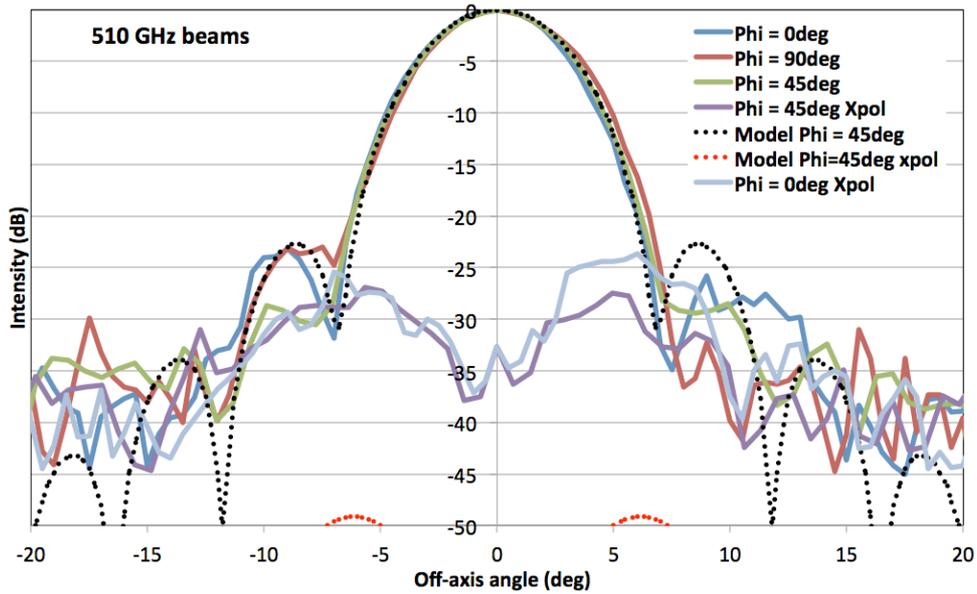

Figure 8. Measured co-polarisation beams cuts for φ=0,45 and 90 deg and cross-polarisation beam cuts for φ=0 and 45 deg at 510 GHz. To be compared with the model (dashed line). An ideal horn would have no cross-polarisation for φ=0 deg.

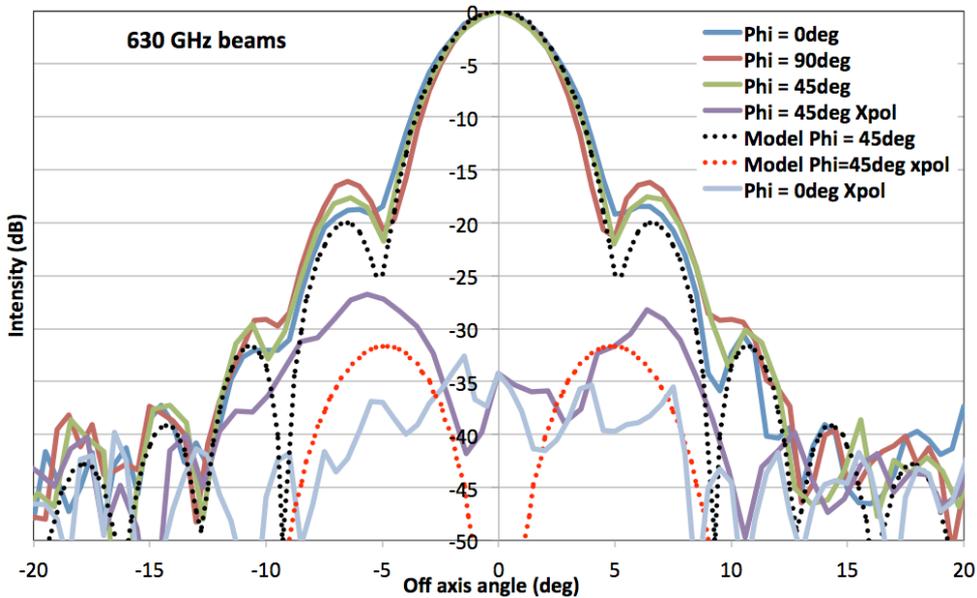

Figure 9. Measured co-polarisation beams cuts for φ=0,45 and 90 deg and cross-polarisation beam cuts for φ=0 and 45 deg at 630 GHz. To be compared with the model (dashed line). An ideal horn would have no cross-polarisation for φ=0 deg.

A corrugated horn manufactured without any error (ideal) will show no cross-polarisation for ϕ = 0 or 90 deg beam cuts. For ϕ = 0, the experimental data show a peak cross-polarisation ranging between -25 and -35 dB, depending on the frequency. This deviation is most likely due to a combination of mechanical imperfections (mainly alignment errors at the waveguide flange and the step transition) and measurement limits that creates a co-polarisation leakage in the cross-polarisation beam. This limits the quality of the ϕ = 45 deg cross-polarisation beam cut when it is expected to be low (at 510 GHz for instance). However, when the predicted cross-polarisation is higher (720 GHz) and becomes the dominant source, the measured ϕ = 45 deg cross-polarisation beam cut follows the same trend (double peaks and dip in the middle). The integral of the cross-polarisation across the main beam will show that we obtain a maximum value below -25 dB across the band.

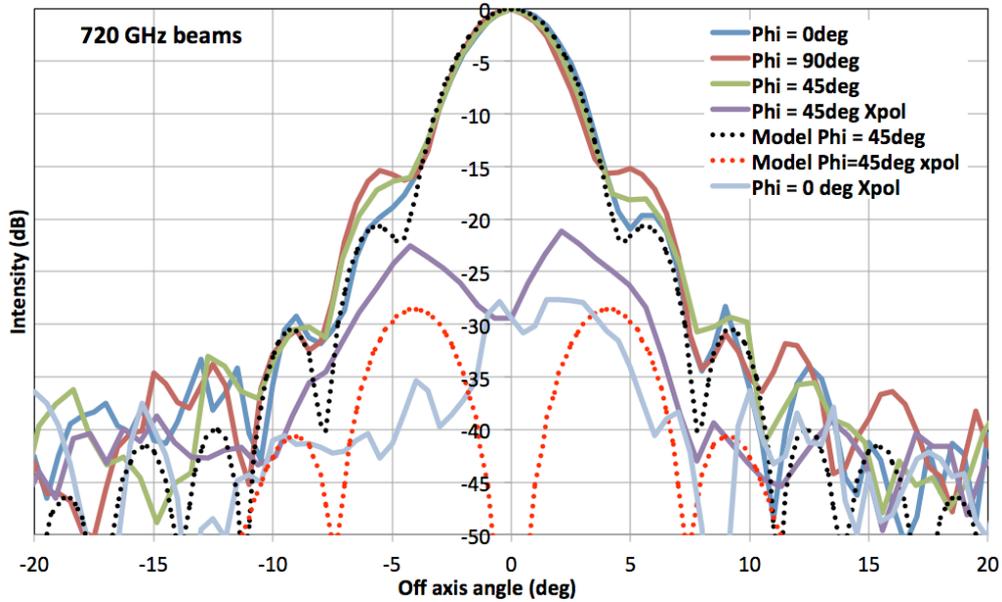

Figure 10. Measured co-polarisation beams cuts for ϕ=0,45 and 90 deg and cross-polarisation beam cuts for ϕ=0 and 45 deg at 720 GHz. To be compared with the model (dashed line). An ideal horn would have no cross-polarisation for ϕ=0 deg.

## 4.4 Directivity

From the beam measurements, the directivity of the horn with frequency is compared with the model in Figure 11. Again this plot shows a good agreement between the model and the experimental results on the manufactured prototype, summarising the remarkable main beam performance of this first high-frequency prototype.

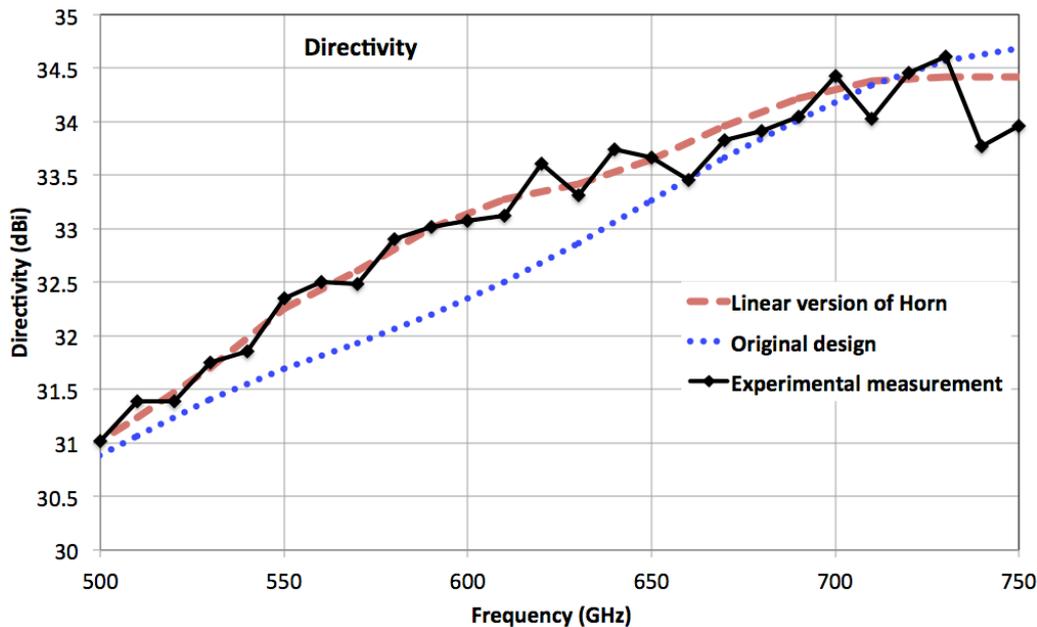

Figure 11. Directivity comparison between the original (dotted blue) and linear (dashed red) models of the horn, and the experimental data deduced from the beam measurements (solid black).

## 5. CONCLUSION

A WR1.5 corrugated feedhorn has been successfully manufactured with the stacked rings technology developed by SWISSto12. While the sidelobes are in general a few dBs higher than predicted, the main beam properties, including the cross-polarisation and directivity, are already good enough for this horn to be use in most applications, even marginally reaching the specifications for CMB observation applications for which horns must meet the most stringent requirements. Nevertheless, an improved version of the presented horn is currently under development. Improvements include tighter mechanical tolerances for the positioning of the crucial step transition and the use of anti-cocking flanges.